\documentclass[AMA,STIX1COL]{WileyNJD-v2}

\usepackage{graphicx}
\articletype{Article Type}%
\usepackage{pgfplots}
\pgfplotsset{width=10cm,compat=1.9}
\received{}
\revised{}
\accepted{}
\usepackage{tikz}
\usetikzlibrary{shapes,arrows}
\begin{document}

\title{Uncertainty Decomposition of Quantum Networks in SLH Framework}

\author[1]{Peyman Azodi*}
\author[1]{Peyman Setoodeh}
\author[1]{Alireza Khayatian }
\author[1]{Mohammad Hassan Asemani}

\authormark{AUTHOR ONE \textsc{et al}}

\address[1]{\orgdiv{School of Electrical and Computer Engineering}, \orgname{Shiraz University}, \orgaddress{\state{Shiraz}, \country{Iran}}}

\corres{*Peyman Azodi. \email{p$\_$azodi@shirazu.ac.ir}}

\presentaddress{Adaptive and Robust control Lab, Shiraz University, 
Shiraz, Iran.}

\abstract[Summary]{This paper presents a systematic method to decompose uncertain linear  quantum input-output networks into uncertain and nominal sub-networks, when uncertainties are defined in SLH representation. To this aim, two decomposition theorems are stated, which show how an uncertain quantum network can be decomposed into nominal and uncertain sub-networks in cascaded connection and how uncertainties can be translated from SLH parameters into state-space parameters. As a potential application of the proposed decomposition scheme, robust stability analysis of uncertain quantum networks is briefly introduced. The proposed uncertainty decomposition theorems take account of uncertainties in all three parameters of a quantum network and bridge the gap between SLH modeling and state-space robust analysis theory for linear quantum networks.}


\jnlcitation{\cname{%
\author{p}, 
\author{p}, 
\author{p}, 
\author{p}, and 
\author{p}} (\cyear{2017}), 
\ctitle{Uncertainty decomposition}, \cjournal{journal}, \cvol{2017;00:1--6}.}

\maketitle

\section{INTRODUCTION}\label{sec1}

\par Developments in quantum technology have led to new ways of engineering our world. Quantum physics has shed light on many dark corners of science and provided justification for a number of counter-intuitive experimental results. It has given rise to a new way of modeling the nature. This modeling scheme results in new ways of thinking about and interpreting of physical phenomena, which in turn, lead to novel approaches to computation and communication. Building on these methods, quantum technology has opened a unique window of opportunity for industrial advancement, and therefore, has attracted the attention of many researchers.
\par Large scale quantum networks represent an important class of quantum systems. This class of quantum systems is assembled through composition of localized optical components that interact with each other via bosonic quantum fields. Each component is characterized by its input-output dynamic behaviour, which is modeled with the triplet $(S,L,H)$\cite{gough2009series,gough2010squeezing,combes2017slh}. Throughout this paper, these systems are called "Quantum Input-Output Networks". SLH formalism is a suitable mathematical tool to model quantum input-output networks. It is also used in modeling superconducting circuits \cite{muller2018passive,cook2018input}, silicon photonic networks \cite{sarovar2016silicon}, and other quantum computing frameworks. Linear quantum networks can also be modeled in state-space realization, using Quantum Stochastic Differential Equations (QSDE) \cite{belavkin1992quantum}.

\par Linear quantum input-output networks, as large-scale quantum systems, are highly affected by environmental disturbances and uncertainties. Quantum and classical noises, assembling limitations, and frequency detuning are the most important sources of uncertainty in linear quantum networks. The impact that these uncertainties may have one the stability and performance of quantum networks highlights the importance of  studying uncertainties in such systems. 
\par Linear quantum input-output network framework is ideal to study and implement quantum coherent feedback control since it is easy to assemble, measure, and control coherently\cite{combes2017slh}. Despite these important positive aspects, uncertainties may have destructive impacts on the stability and performance of these systems. Robust coherent control has been developed to study and facilitate the robust performance of these systems in recent years. As mentioned, linear quantum networks are easily modeled in SLH formalism, but the existing robust control theories are applicable only when these systems are modeled in state-space realization; thus, one of the main obstacles in robust coherent control of linear quantum networks is to transform uncertainties from SLH formalism into state-space realization. In this paper, uncertainty decomposition scheme is introduced to transform uncertainties in all three parameters of a linear quantum network from SLH formalism into state-space realization of these systems in two steps. In the first step, an uncertain linear quantum network is decomposed into two (uncertain and nominal) sub-networks in SLH formalism, and in the second step, methods are proposed to transform uncertainties in SLH parameters of the sub-networks into uncertainties in state-space realization of the original system.
\par In the literature, modeling linear quantum networks in both SLH formalism and state-space realization is well-introduced and studied \cite{combes2017slh, mabuchi2005principles,dong2010quantum,gough2009series,gough2010squeezing}. After the introduction of state-space realization for linear input-output quantum networks, robust analysis techniques, mainly $H^\infty$ and risk-sensitive analysis and controller synthesis methods are applied to these systems \cite{james2008h,doi:10.1098/rsta.2011.0527,d2006stability,yamamoto2006robust}. Also, new robust analysis, controller synthesis, and filtering algorithms are developed for different classes of such systems \cite{roy2016robust,wang2017robust,maalouf2017finite}. In these proposed robust analysis methods, uncertainties are introduced in the state-space realization of linear quantum networks. In \cite{doi:10.1098/rsta.2011.0527} authors introduced additive uncertainty in the Hamiltonian part of SLH formalism and proposed a method to evaluate robust stability of the quantum system. In a recently published article, \cite{xiang2017coherent} authors introduced uncertainties in both Hamiltonian and coupling operator, and proposed a coherent robust controller for the system. So far, presence of uncertainties in all three parameters of a linear quantum system has not been thoroughly studied. The proposed decomposition theorems in this paper take uncertainties in all three parameters of SLH realization of quantum networks into account.

\par This paper is organized as follows. Section \ref{s1} provides an introduction to linear quantum input-output networks and their modeling. In section \ref{sec2}, uncertain linear quantum systems in SLH realization are described. The proposed uncertainty modeling scheme is introduced and justified in \ref{311} and \ref{3122}, respectively. In section \ref{444}, the main results of this paper are presented. In this section, two decomposition theorems are presented, which include systematic decomposition algorithms in both SLH and state-space representations, and how they can be transformed to each other. In subsection \ref{4441}, a robust stability theorem is proposed based on uncertainty decomposition as an application of the decomposition theorems. Then, an illustrative example is presented in section \ref{s3} to validate the systematic procedures of the proposed algorithms. The paper concludes in Section \ref{s4}. The Appendix provides two lemmas used in the proof of theorems.

\section{Introductory review of SLH formalism}\label{s1}
In this section, SLH formalism for modeling linear quantum input-output networks is briefly reviewed in order to help readers to easily follow the subsequent derivations. Further details on the algebra and applications can be found in the recently published survey \cite{combes2017slh} or relevant references \cite{gough2010squeezing,gough2009series,PhysRevA.31.3761}.
\subsection{Notation}
\par In what follows, the notation used for modeling a linear quantum network in SLH formalism is explained.
\begin{itemize}
    \item { For a $2m \times 2n$ matrix $Z=({z_{ij}})$, let us define ${Z^\# } \triangleq  (z_{ij}^*)$, ${Z^T} \triangleq (z_{ji})$, and  ${Z^\dag } \triangleq  ( z_{ji}^*)$, with $z_{ij}$ being a complex number , and ${Z^\flat } \triangleq {J_m}{Z^\dag }{J_n}$, where
$${J_n} \triangleq \left( {\begin{array}{*{20}{c}}
  {{I_n}}&0 \\ 
  0&{ - {I_n}} 
\end{array}} \right).$$
}
\item { Let  $X(t) = {\left( {{x_1}(t),{x_2}(t),...,{x_n}(t)} \right)^{\rm T}}$ be a vector of operators for all different degrees of freedom of the system. The doubled-up notation of $X(t)$ is defined as:
$$\mathop X\limits^ \cup   \triangleq \left( {\begin{array}{*{20}{c}}
  X \\ 
  {{X^\# }} 
\end{array}} \right).$$}

\item{ Consider a linear combination of the system states as follows:
$$Y = {E^ - }X + {E^ + }{X^\# };{E^ - }\& {E^ + } \in {\mathbb{C}^{m \times n}},$$
in the doubled-up form, it is written as:
$$\mathop Y\limits^ \cup   = \Delta ({E^ - },{E^ + })\mathop X\limits^ \cup,  $$
with
\[\Delta ({E^ - },{E^ + }) \triangleq \left( {\begin{array}{*{20}{c}}
  {{E^ - }}&{{E^ + }} \\ 
  {{E^ + }^\# }&{{E^ - }^\# } 
\end{array}} \right).\]}
    
\end{itemize}

\subsection{Elements of SLH formalism}
\par A general Multi-Input-Multi-Output (MIMO) linear quantum network contains active components (beam splitter and cavity) and passive components (amplifiers). Each of these components is characterized by three elements $(S,L,H)$. Also, there are rules for obtaining the SLH model of the network from its components in cascaded, concatenated, and feedback form wiring \cite{combes2017slh, gough2009series}.
\par In this formalism, $H$ characterizes internal energy dynamics of the system and $ (S,L)$  specify the interface of the system to external bosonic field channels:
\begin{itemize}
\item{Scattering matrix, $S \in {\mathbb{C}^{m \times m}}$ , is a unitary matrix (${S^\dag }S = S{S^\dag } = I$ ), which corresponds to the input-output fields static relation.}
\item{Coupling operator, $L = C\mathop X\limits^ \cup   = \left( {\begin{array}{*{20}{c}}
  {{C^ - }},&{{C^ + }} 
\end{array}} \right)\mathop X\limits^ \cup  $ , with $C \in {\mathbb{C}^{m \times 2n}}$,  describes how input and output fields interact with the components' internal dynamics.}
\item{Hamiltonian, $H$, is Hermitian and characterizes the intrinsic (not influenced by input fields) time evolution of system states in Heisenberg picture.}
\end{itemize}
\subsection{Input-Output fields}\label{Ioo}
\par Each $m-$input, $m-$output quantum network (or component as a sub-network) is driven by $m$ independent bosonic annihilation and creation quantum field operators, ${\rm A}_i^{in}(t) \text{ and } {{\rm A}^\dag}_i^{in}(t); i:1,...,m$, defined on separate Fock spaces ${F_i}$. Output fields are denoted by ${\rm A}_i^{out}(t) \text{ and } {{\rm A}^\dag}_i^{out}(t); i:1,...,m$, correspondingly. 
\par These field operators are adapted quantum stochastic processes with the following quantum Ito products:
$$\begin{gathered}
  d{\rm A}_i^{}(t)d{{\rm A}^\dag}_j^{}(t) = {\delta _{ij}}dt, \\ 
  d{{\rm A}^\dag}_i^{}(t)d{{\rm A}^\dag}_j^{}(t) = 0, \\ 
  d{\rm A}_i^{}(t)d{\rm A}_j^{}(t) = 0, \\ 
  d{{\rm A}^\dag}_i^{}(t)d{\rm A}_j^{}(t) = 0. \\ 
\end{gathered} $$
\par The vector of input (output) fields is denoted by:
$${\rm A^{in(out)}}(t) = {\left( {{\rm A}_1^{}(t),{\rm A}_2^{}(t),...,{\rm A}_m^{}(t)} \right)^{\rm T}}.$$

\subsection {Open quantum harmonic oscillator modeling} \label{openquan}

\par The key ingredient in modeling quantum optical elements is creation and annihilation of photonic (and phononic) quanta in each independent frequency modes of the system, modeled by non-Hermitian operators. These operators arise due to second-quantizing the electromagnetic fields. In semi-classical regime, coherent states appear as the eigenkets of the creation operators, which are frequently used to model laser-driven optical fields. Creation and annihilation operators (and less frequently quadrature operators, which are real and imaginary parts of the annihilation operator) are used as system's states (in modeling passive optical components, creation operators are usually neglected).
\par For a system with $n$ different frequency modes (also called degrees of freedom), the doubled up state vector $\mathop X \limits^ \cup (t)$ contains $n$  independent annihilation operators ${a_i(t)}$, and creation operators $a_i^\dag (t)$, with the following canonical commutation relations: 
$$\begin{gathered}
  \left[ {{a_i}(t),a_j^\dag (t)} \right] = {\delta _{ij}} \\ 
  \left[ {{a_i}(t),{a_j}(t)} \right] = \left[ {a_i^\dag (t),a_j^\dag (t)} \right] =0 . \\ 
\end{gathered} $$

\par A general form for the Hamiltonian operator, taking into account both passive (photon number preserving) and active components, is the following quadratic form (if neglecting vacuum energy and putting $\hbar=1$):
\begin{equation} \label{a}
H = \sum\limits_{i,j = 1}^n {\left( {\omega _{ij}^{-} a_i^{\dag}{a_j} + \frac{1}{2}\omega _{ij}^ + a_i^{\dag}a_j^{\dag} + \frac{1}{2}\omega {{_{ij}^ + }^*}a_i^{}a_j^{}} \right)},
\end{equation}
with $\omega _{ij}^ - , \omega _{ij}^ +  \in \mathbb{C}$.
\\ In correspondence with this quadratic form of Hamiltonian (\ref{a}), let us define:
$${\Omega _ - } = (\omega _{ij}^ - ),{\Omega _ + } = (\omega _{ij}^ + ) \in {\mathbb{C}^{n \times n}},$$
$$ - i\mathop \Omega \limits^ \sim   =  - \Delta (i{\Omega _ - },i{\Omega _ + })$$
$$H \sim \tilde \Omega. $$
where $ \sim$ denotes the correspondence between the Hamiltonian and its doubled-up form $\tilde \Omega$.

\subsection{Cascaded connection of components}
Any two components in SLH formalism, up to some limitations imposed by the number of fields, can interconnect with each other through their input-output fields. These connections are typically cascaded, concatenated, feedback-form, padding-form, and channel permuting connections or other combinations of these forms. There are rules for deriving the SLH form of the composite component from its sub-components\cite{combes2017slh,gough2009series,gough2010squeezing}. In this paper, cascaded connection is investigated. In the following sections, algebra of cascaded components is used to derive the uncertainty decomposition rules. 
\par Let  ${G_1} = ({S_1},{L_1},{H_1})$  and  ${G_2} = ({S_2},{L_2},{H_2})$ be two optical components. In the cascaded connection,  $G = {G_2} \triangleleft {G_1}$, shown in figure \ref{figcas}, the output fields of ${G_1}$ are fed into the input fields of ${G_2}$.
\tikzstyle{block} = [draw, rectangle, 
    minimum height=3em, minimum width=6em]

\tikzstyle{input} = [coordinate]
\tikzstyle{output} = [coordinate]
\tikzstyle{pinstyle} = [pin edge={to-,thin,black}]

\begin{figure}[t]
\begin{center}
\begin{tikzpicture}[auto, node distance=2cm,>=latex']
    \node [input, name=input] {};
    \node [block, right of=input,node distance=2cm] (controller) {$G_1$};
    \node [block, right of=controller, node distance=3.2cm] (system) {$G_2$};
    \node [output, right of=system, node distance=2cm] (output) {};
    \draw [->] (input) -- node[name=u] {} (controller);
    \draw [->] (controller) -- node[name=u] {} (system);
    \draw [->] (system) -- node[name=u] {} (output);
\end{tikzpicture}
\end{center}
\caption{Cascaded connection.\label{figcas}}
\end{figure}
One may write the model parameters of $G = (S,L,H)$  as follows: 
\begin{equation}\label{12}
G = \left( {{S_2}{S_1},{L_2} + {S_2}{L_1},{H_1} + {H_2} + \operatorname{Im} \left( {L_2^\dag {S_2}{L_1}} \right)} \right).
\end{equation}
\subsection{State-space realization from SLH form}\label{sec}
\par Linear input-output quantum networks can also be described in time domain by state-space realization, $G \equiv (\tilde A,\tilde B,\tilde C,\tilde D)$. 
Consider the quantum network $G \equiv (S,L,H)$, in SLH formalism, described by state vector $X (t)$, as discussed in subsection \ref{openquan}, which is coupled to input and output field vectors ${A^{in}}(t)$  and ${A^{out}}(t)$, as discussed in subsection \ref{Ioo}. Stochastic state-space realization of the system is as follows\cite{gough2010squeezing}:
\[\begin{gathered}
  d\mathop X\limits^ \cup  (t) = \tilde A\mathop X\limits^ \cup  dt + \tilde Bd{\mathop A\limits^ \cup  }^{in}(t), \hfill \\
  d{\mathop A\limits^ \cup  }^{out}(t) = \tilde C\mathop {\mathop X\limits^ \cup  }\limits^{} (t)dt + \tilde Dd{\mathop A\limits^ \cup  }^{in}(t) \hfill, \\ 
\end{gathered} \]
where
\begin{equation}\label{e33}
\begin{gathered}
  \tilde A =  - \frac{1}{2}{{\tilde C}^{\flat} }\tilde C - i\tilde \Omega ,\hspace{3mm} \tilde B =  - {{\tilde C}^{\flat} }\tilde D,\hspace{3mm} \tilde C = \Delta ({C^ - },{C^ + }),\hspace{3mm} \tilde D = \Delta (S,0), \\ 
  H \sim \tilde \Omega.  \\ 
\end{gathered}
\end{equation}
\par State-space realization is a useful and insightful representation of a system in the sense that many dynamic properties can be extracted from parameters $(\tilde A,\tilde B,\tilde C,\tilde D)$. One of the most important applications of this realization is the so-called $H^{\infty}$ robust analysis and controller synthesis. \par Based on this brief review of SLH and state-space realization of linear quantum networks, in the rest of this paper, as the first step for robust analysis and controller synthesis of uncertain linear quantum networks, rules for decomposing an uncertain linear quantum network into nominal and uncertain parts are presented.

\section{UNCERTAIN LINEAR QUANTUM NETWORKS}\label{sec2}
\par Parameter uncertainty is prevalent in systems theory and it can affect the stability and performance of the system. For classical systems, one class of uncertainties is due to the lack of information about the system, engineering limitations, and aging. The second class of uncertainties is due to simplifying methods used for modeling a system; for example, order-reduced systems in which just a few dominant modes of a system are considered, or linearized non-linear systems, in which the approximated linear behavior of an intrinsically nonlinear system is considered. In quantum case, there is another source of uncertainty, which is due to the nature of the system. This class of uncertainty appears when an observer wants to simultaneously measure two non-commuting observables. In this case, based on the postulates of quantum mechanics, there exists a positive lower bound on the uncertainties of the measurement of the two quantities. In this paper, we do not model this class of uncertainty explicitly, since it has already been considered in deriving creation and annihilation operator algebras implicitly.  Also, as any other physical systems, in dealing with quantum systems, taking parameter uncertainties into account deserves special attention. In these systems, other factors such as detuning parameters and implementation limits are also sources of uncertainty. In this section, a general uncertainty modeling scheme is proposed. This scheme, as will be discussed in detail, is based on the nature of the parameters and is inspired by the way the parameters contribute in modeling a quantum input-output system.
\subsection{Uncertainty modeling}\label{311}
\par Regarding a general uncertainty form, given an uncertain quantum network, $G \equiv (S,L,H)$, the scattering matrix, $S$, is assumed to be perturbed in post multiplicative form, $S = {S_n}\Delta S$, where ${S_n}$  is the nominal parameter and $\Delta S$  is the perturbation part. The coupling matrix, $L$, and the system's Hamiltonian, $H$, are assumed to be perturbed in additive form, $L = {L_n} + \Delta L$ and $H = {H_n} + \Delta H$, respectively, where ${L_n}$ and ${H_n}$ are the nominal coupling operator and Hamiltonian, and $\Delta L$  and  $\Delta H$ are their additive perturbation parts, respectively. Based on the prior knowledge about the system and the sources of uncertainty, we consider three uncertainty sets, $\overline {\Delta S} $  , $\overline {\Delta L} $  and$\overline {\Delta H}$, which include $\Delta S$, $\Delta L$ and $\Delta H$, respectively and reflect their properties. For instance, $\overline {\Delta L}$  may be defined as $\overline {\Delta L}  = \left\{ {\left( {{\delta _c},0} \right)\overset{\lower0.5em\hbox{$\smash{\scriptscriptstyle\smile}$}}{X} \left| {{\delta _c} \in {\mathbb{C}^{m \times n}},{{\left| {{\delta _c}} \right|}_\infty } \leqslant 1} \right.} \right\}$.
\par These uncertainty sets usually differ for any two different uncertain linear quantum systems regarding their differences in structure and the sources of uncertainty. Despite the differences, these uncertainty sets must necessarily satisfy the following conditions:
\begin{boxtext}

\begin{itemize}
\item$\forall \Delta S \in \overline {\Delta S} ;$$\Delta S$ is unitary. 
\item $\overline {\Delta L}  \subset \left\{ {\left( {\delta _c^ - ,\delta _c^ + } \right)\mathop X\limits^ \cup  \left| {\delta _c^ - \& \delta _c^ +  \in {\mathbb{C}^{m \times n}}} \right.} \right\}$.
\item $\forall \Delta H \in \overline {\Delta H} ;$$\Delta H$ has the mathematical form in (\ref{a}).
\end{itemize}
\end{boxtext}

Uncertainty sets $\overline {\Delta S} $,  $\overline {\Delta L} $ ,and $\overline {\Delta H} $, together with the nominal parameters ${S_n}$, ${L_n}$, and ${H_n}$, form the set of admissible systems $\overline G $, which consists of all the admissible linear quantum systems of the form $\left( {{S_n}\Delta S,{L_n} + \Delta L,{H_n} + \Delta H} \right)$, where the perturbation parts $\Delta S$, $\Delta L$ , and $\Delta H$ belong to the sets, $\overline {\Delta S} $, $\overline {\Delta L} $, and $\overline {\Delta H} $, respectively:
\begin{equation}
\overline G  = \left\{ \begin{gathered}
  \left( {{S_n}\Delta S,{L_n} + \Delta L,{H_n} + \Delta H} \right)\left| {\Delta S \in \overline {\Delta S} } \right., \hfill 
  \Delta L \in \overline {\Delta L} ,\Delta H \in \overline {\Delta H}  \hfill
\end{gathered}  \right\}.
\end{equation}
\subsection{Generality of the decomposition}\label{3122}

\par The framework for uncertainty modeling, which is considered here, was chosen based on experimental applications and nature of parameters. For Hamiltonian operator, the most frequent uncertainties in experimental quantum optics arise in additive manner and this is due to the simple reason that the total Hamiltonian including the influence of different interactions with commutative Hamiltonians is evaluated by adding up the Hamiltonian of each single interaction. Also, based on (\ref{e33}), additive perturbation of the Hamiltonian operator simply induces an additive part to the state matrix $\tilde A$. Hence, additive perturbation of the Hamiltonian operator is very common in quantum systems. As a result, quantum perturbation theory studies the effect of additive perturbations of Hamiltonian operator. Therefore, choosing additive perturbation for Hamiltonian operator seems rational based on both experimental and natural reasons. Also, in \cite{yamamoto2006robust,james2008h,petersen2012robust} additive form of Hamiltonian perturbation was considered.
\par Although uncertainties in coupling operator is less trivial than in Hamiltonian operator, readers may find it absolutely reasonable to consider it in additive manner, based on the following justifications. From one viewpoint, quantum Langevin equation dictates that any influence in the state evolution of the system due to couplings add up linearly to the input fields \cite{gardiner1985input}. This is a consequence of first Markov approximation and that is the underlying reason that the coupling operator of a cascaded system is calculated by addition of the single coupling operators (up to a multiplication by the scattering matrix). Also, based on (\ref{e33}), and as will be shown in the subsequent sections, additive perturbation of coupling operator, will add additive perturbation terms to the state, input and output matrices, which would be suitable for robust stability analysis. Also, in \cite{yamamoto2006robust}, additive uncertainty form was considered for coupling operator in order to design a robust quantum observer. 
\par The multiplicative uncertainty form for the scattering matrix is chosen based on both its multiplicative and unitary nature, which gives a better explanation for uncertainty in a unitary matrix and scattering phenomenon than the additive uncertainty form. The scattering matrix describes how input beams are scattered into output beams, and the unitarity property is because of the non-dissipative nature of an ideal component. The main reason for choosing multiplicative perturbation is that the multiplicative perturbation is compatible with the propagative nature of the scattering matrix. The other reason is due to energy-preserving reason that the unitary multiplicative perturbation preserves the unitarity of the scattering matrix.
\par It should be noted that the most relevant way to model dissipation in elements or channels of a quantum network is to use an extra pure-scattering element (beam-splitter) as in \cite{motzoi2015continuous,motzoi2016backaction} or multiplicative perturbation of the scattering matrix of the element. Also, this approach to model dissipation is worthwhile as the exact amount of dissipation is usually unknown and time-varying, and in this scheme, it is enough to consider an upper-bound on the dissipation in the uncertainty set $\overline {\Delta S} $.   

\section{decomposition theorems}\label{444}
\par In the following, a decomposition theorem is presented. The goal of this decomposition theorem is to decompose an arbitrarily perturbed linear quantum input-output network ($G \in \overline G $) into two subsystems.
\begin{figure}[t]

\begin{center}
\hspace{1cm}
\begin{tikzpicture}[auto, node distance=2cm,>=latex']
    \node [input, name=input] {};
    \node [block, right of=input,node distance=2cm] (controller) {$G$};
    
    \draw [->] (input) -- node[name=u] {} (controller);
    \draw [->] (controller) -- node[name=u] {} (system);
\end{tikzpicture}
$\hspace{1cm} \xrightarrow{{Decomposition}} \hspace{1cm}$
\begin{tikzpicture}[auto, node distance=2cm,>=latex']
    \node [input, name=input] {};
    \node [block, right of=input,node distance=2cm] (controller) {$\Delta$};
    \node [block, right of=controller,node distance=3.2cm] (system) {$G_n$};
    \node [output, right of=system, node distance=2cm] (output) {};
    \draw [->] (input) -- node[name=u] {} (controller);
    \draw [->] (controller) -- node[name=u] {} (system);
    \draw [->] (system) -- node[name=u] {} (output);
\end{tikzpicture}
\end{center}
\caption{Decomposition of an uncertain linear quantum network into two sub-networks.\label{fig1}}

\end{figure}
\begin{theorem}\label{th1}
Every uncertain linear quantum input-output network $G \in \overline G $, with uncertainty decomposition structure introduced in section \ref{311} can be decomposed into two linear quantum sub-networks $\Delta $ and ${G_n}$, in a cascaded manner, $G \equiv {G_n} \triangleleft \Delta $, as shown in figure \ref{fig1}, where the underlying sub-networks possess the following parameters:
\begin{equation}\label{222}
\begin{gathered}
  G \equiv ({S_n}\Delta S,{L_n} + \Delta L,{H_n} + \Delta H), \\ 
  {G_n} \equiv ({S_n},{L_n},{H_n}), \\ 
  \Delta  \equiv (\Delta S,{S_n}^\dag \Delta L,\Delta H - \operatorname{Im} (L_n^\dag \Delta L)). \\ 
\end{gathered} 
\end{equation}
\end{theorem}
\begin{proof}
This decomposition consists of two blocks connected in a cascaded manner; each block is realized by the parameters in (\ref{222}). Using (\ref{12}), the composition system parameters are obtained from individual systems ($G \equiv {G_n} \triangleleft \Delta $ ) as:
$$G \equiv ({S_n}\Delta S,L_n^{} + {S_n}{S_n}^\dag \Delta L,{H_n} + \Delta H - \operatorname{Im} (L_n^\dag \Delta L) + \operatorname{Im} (L_n^\dag {S_n}{S_n}^\dag \Delta L))$$
The fact that ${S_n}$  is unitary (${S_n}S_n^\dag  = I$ ), implies that:
$$G \equiv ({S_n}\Delta S,L_n^{} + \Delta L,{H_n} + \Delta H)$$
which is equivalent to the main system $G$.
\end{proof}
By this decomposition theorem, an uncertain linear quantum system is decomposed into two subsystems, one of which is completely certain and the other one consists of uncertainties. In this way, nominal parameters are completely separated from uncertain parameters. The following points are worth mentioning regarding this theorem.
\begin{remark}\label{r1}
The Proposed decomposition is a non-minimal realization of the original system . Since this decomposition is applied to the uncertain system under consideration, both subsystems share the same state vector.
\end{remark}
\begin{remark}
By this decomposition, the nominal system's response to external fields becomes more apparent.  However, there is no particular relation (i.e., linear combination or multiplication) between the states of the original system, $G$, and the states of the subsystems $\Delta$ and $G_n$. In particular, this decomposition is an artificial decomposition in order to facilitate the procedure of robustness analysis and control.
\end{remark}

In the following, an extended decomposition theorem is presented, which decomposes the state-space parameters of the original uncertain linear quantum system in the form of state-space representation and SLH parameters of its subsystems according to Theorem \ref{th1}. 
Consider the uncertain system, $G$, and subsystems $G_n$ and $\Delta$. Using the methods mentioned in Section \ref{sec}, their state-space representations will be:
\begin{equation}\label{e1}
\begin{gathered}
  G \equiv (\tilde A,\tilde B,\tilde C,\tilde D), \\ 
  {G_n} \equiv ({{\tilde A}_n},{{\tilde B}_n},{{\tilde C}_n},{{\tilde D}_n}), \\ 
  \Delta  \equiv ({{\tilde A}_\delta },{{\tilde B}_\delta },{{\tilde C}_\delta },{{\tilde D}_\delta }). \\ 
\end{gathered} 
\end{equation}
Now we can state the following theorem:
 \begin{theorem}\label{th2}
Consider an uncertain linear quantum input-output network, $G \in \overline G $, decomposed into sub-networks $G_n$ and $\Delta$ ( $G \equiv {G_n} \triangleleft \Delta $) based on the algorithm introduced in Theorem \ref{th1}. The state-space realization of the systems are denoted by the notation introduced in (\ref{e1}). Then, the following relations between the state-space realizations hold:
\begin{enumerate}
\item{The state matrices are related as:}\label{a1}
\begin{equation}\label{a234}
\tilde A = {\tilde A_n} + ({\tilde A_\delta } + \tilde A') = {\tilde A_n} + \Delta \tilde A,
\end{equation}
where:
$$\tilde A' =  - {\tilde C_n}^\flat S_n^\dag {\tilde \delta _c}.$$
Also, the additive perturbation matrix is calculated as follows:
$$\Delta \tilde A =  - \frac{1}{2}{\tilde \delta _c}^\flat {\tilde \delta _c} - i{\tilde \Omega _{\Delta H}} - {\operatorname{Re} _\beta }({\tilde C_n}^\flat S_n^\dag {\tilde \delta _c}),$$
where ${\tilde \delta _c} \triangleq \Delta (\delta _c^ - ,\delta _c^ + )$ and ${\tilde \Omega _{\Delta H}} \sim \Delta H$.
\item{The output and feed-forward matrices are related as:}\label{a2}
\begin{equation}\label{e44}
\begin{gathered}
  \tilde C = {{\tilde C}_n} + {{\tilde C}_\delta }, \\ 
  \tilde D = {{\tilde D}_n}{{\tilde D}_\delta }. \\ 
\end{gathered} 
\end{equation}
\item{The input matrices are related as:}\label{a3}
$$\tilde B = {\tilde B_n}{\tilde D_\delta } - {\tilde C_\delta }^\flat {\tilde D_n}{\tilde D_\delta }.$$
\end{enumerate}
\end{theorem}
\begin{proof}
Consider the following decomposition due to Theorem \ref{th1}:
$$\begin{gathered}
  G \equiv ({S_n}\Delta S,{L_n} + \Delta L,{H_n} + \Delta H), \\ 
  {G_n} \equiv ({S_n},{L_n},{H_n}), \\ 
  \Delta  \equiv (\Delta S,{S_n}^\dag \Delta L,\Delta H - \operatorname{Im} (L_n^\dag \Delta L)), \\ 
\end{gathered} $$
where
$$\begin{gathered}
  {L_n} = {C_n}\mathop X\limits^ \cup   = \left( {\begin{array}{*{20}{c}}
  {{C_n}^ - }&{{C_n}^ + } 
\end{array}} \right)\mathop X\limits^ \cup,   \\ 
  \Delta L = {S_n}^\dag {\delta _C}\mathop X\limits^ \cup   = \left( {{S_n}^\dag \begin{array}{*{20}{c}}
  {{\delta _c}^ - }&{{S_n}^\dag {\delta _c}^ + } 
\end{array}} \right)\mathop X\limits^ \cup,   \\ 
  {H_n} \sim {{\tilde \Omega }_n}, \\ 
  \Delta {H_n} \sim {{\tilde \Omega }_{\Delta H}}. \\ 
\end{gathered} $$
\ref{a1}.\hspace{0.2cm}Using the presented techniques in Section \ref{sec}, the state matrix of $G$ is:
$$\tilde A =  - \frac{1}{2}\left( {{{\mathop {({{\tilde C}_n} + {{\tilde S}_n}^\dag {{\tilde \delta }_C})}\limits^{} }^{_\flat }}\mathop {({{\tilde C}_n} + {{\tilde S}_n}^\dag {{\tilde \delta }_C})}\limits^{} } \right) - i{\tilde \Omega _n} - i{\tilde \Omega _{\Delta H}}.$$
Using the algebraic relations in Lemma \ref{l1}, the first term of $\tilde A$ is decomposed into the following sub-terms:
$$\begin{gathered}
  \tilde A =  - \frac{1}{2}\left( {\tilde C_n^\flat \tilde C_n^{} + \tilde \delta _C^\flat \tilde \delta _C^{} + \tilde C_n^\flat {S_n}^\dag \tilde \delta _C^{} + {{\left( {{S_n}^\dag \tilde \delta _C^{}} \right)}^\flat }\tilde C_n^{}} \right) - i{{\tilde \Omega }_n} - i{{\tilde \Omega }_{\Delta H}} \hfill, \\
  {\text{   }} =  - \frac{1}{2}\left( {\tilde C_n^\flat \tilde C_n^{} + \tilde \delta _C^\flat \tilde \delta _C^{}} \right) - {\operatorname{Re} _\flat }(\tilde C_n^\flat {S_n}^\dag \tilde \delta _C^{}) - i{{\tilde \Omega }_n} - i{{\tilde \Omega }_{\Delta H}} \hfill. \\ 
\end{gathered} $$
Using similar calculations, the state matrix ${\tilde A_n}$  will be:
$${\tilde A_n} =  - \frac{1}{2}\tilde C_n^\flat \tilde C_n^{} - i{\tilde \Omega _n}.$$
In order to derive ${\tilde A_\delta }$ , by Lemma \ref{l2}, the perturbation term, $ - \operatorname{Im} (L_n^\dag \Delta L)$, in the Hamiltonian of $\Delta $, induces the term $i{\operatorname{Im} _\flat }(\tilde C_n^\flat {S_n}^\dag \tilde \delta _C^{})$  to${\tilde A_\delta }$:
$${\tilde A_\delta } =  - \frac{1}{2}\tilde \delta _C^\flat \tilde \delta _C^{} - i{\tilde \Omega _{\Delta H}} + i{\operatorname{Im} _\flat }(\tilde C_n^\flat {S_n}^\dag \tilde \delta _C^{}).$$
Considering (\ref{e1}), $\tilde A'$ is written as:
$$\tilde A' = \tilde A - {\tilde A_n} - {\tilde A_\delta } =  - {\operatorname{Re} _\flat }(\tilde C_n^\flat {S_n}^\dag \tilde \delta _C^{}) - i{\operatorname{Im} _\flat }(\tilde C_n^\flat {S_n}^\dag \tilde \delta _C^{}) =  - \tilde C_n^\flat {S_n}^\dag \tilde \delta _C^{}.$$
Also, $\Delta A$  is obtained as follows:
$$\Delta \tilde A = \tilde A' + {\tilde A_\delta }{\text{  = }} - \frac{1}{2}\tilde \delta _C^\flat \tilde \delta _C^{} - i{\tilde \Omega _{\Delta H}} + {\operatorname{Re} _\flat }(\tilde C_n^\flat {S_n}^\dag \tilde \delta _C^{}).$$
\ref{a2}.\hspace{0.2cm} From Lemma \ref{l1} and the fact that coupling operator is perturbed by additive perturbation, we have:
$$\tilde C = \Delta ({C^ - } + \delta _c^ - ,{C^ + } + \delta _c^ + ) = \Delta ({C^ - },{C^ + }) + \Delta (\delta _c^ - ,\delta _c^ + ) = {\tilde C_n} + {\tilde C_\delta }.$$
Also, the scattering matrices are perturbed in multiplicative form; so, using (\ref{e33}) gives:
$$\tilde D = \Delta ({S_n}\Delta S,0) = \Delta ({S_n},0)\Delta (\Delta S,0) = {\tilde D_n}{\tilde D_\delta }.$$
\ref{a3}.\hspace{0.2cm} Using (\ref{e33}), (\ref{e44}), and Lemma \ref{l1}, $\tilde B$ can be rewritten as:
$$\tilde B =  - {\tilde C^\flat }\tilde D =  - {({\tilde C_n} + {\tilde C_\delta })^\flat }{\tilde D_n}{\tilde D_\delta } =  - ({\tilde C_n}^\flat  + {\tilde C_\delta }^\flat ){\tilde D_n}{\tilde D_\delta } = {\tilde B_n}{\tilde D_\delta } - {\tilde C_\delta }^\flat {\tilde D_n}{\tilde D_\delta }.$$
\end{proof}
According to Theorem \ref{th2}, the state-space parameters of the uncertain linear quantum input-output network is related to the parameters of its subsystems. The state matrix of the uncertain subsystem, ${\tilde A_\delta }$, together with an additional matrix $\tilde A'$ will play the role of an additive perturbation to the nominal subsystem. This statement shows how these two subsystems are connected to each other and contribute to the dynamics of the original uncertain system.
In the next subsection, the extended decomposition theorem is used to derive a robust stability theorem for linear quantum input-output systems as a potential application of the proposed decomposition scheme.

\subsection{Robust stability analysis based on uncertainty decomposition}\label{4441}
In \cite{azodi2016robustly}, authors used the uncertainty decomposition algorithm and proposed two different approaches to analyze robust stability of uncertain linear quantum input-output networks. In this subsection, one of the main results of this robust stability theory is briefly presented in order to illustrate how uncertainty decomposition is applicable in robust analysis of quantum input-output networks.
\begin{definition}
An uncertain linear quantum input-output network, $G \equiv (\tilde A,\tilde B,\tilde C,\tilde D) \in \Bar{G}$, is said to be \underline{robustly stable}, if $\tilde A$ remains Hurwitz stable (negative real part for all eigenvalues) for all $G \in \Bar{G}$.
\end{definition}

Also, assume that the following norm-bounded condition holds on the additive perturbation matrix $\Delta \tilde A$:
\begin{equation}\label{1212}
    \Delta \tilde A^{\dag}\Delta \tilde A\leqslant \eta ^2 I,
\end{equation}
where $\eta$ is a known positive constant. Now, the following robust stability theorem can be stated.
\begin{theorem}\label{Th33}
Assume that the linear quantum input-output network, $G \in \Bar{G}$, is decomposed in the sense described in Theorem \ref{th1} as $G \equiv (\tilde A ={\tilde A_n} + \Delta \tilde A,\tilde B,\tilde C,\tilde D)$ . Also, assume that ${\tilde A_n}$ is Hurwitz stable and $\Delta \tilde A$ satisfies the norm-bounded condition in (\ref{1212}). Then, $G$ is robustly stable if the following condition holds:
\begin{equation}\label{322}
    \eta < \underset{\omega}{\inf} \underline{\sigma}(i\omega I- \tilde A_n ),
\end{equation}
where $\underline{\sigma }(.)$ denotes the smallest singular value. 
\end{theorem}
It is also important to note that in this theorem, for simplicity, the uncertainty structure of the additive perturbation matrix $\Delta \tilde A$ is not taken into account and the $L_2$-norm of this matrix ($\eta$) plays the main role. Thus, the stability region predicted by this theorem might be conservative. However, it gives a very handy tool for robust stability analysis, even for complicated quantum networks, because as shown in \cite{azodi2016robustly}, both sides of (\ref{322}) can be easily calculated using simple Linear Matrix Inequality (LMI) techniques. Using the proposed uncertainty decomposition, other robust stability theorems can be obtained.
\section{ILLUSTRATIVE EXAMPLE}\label{s3}
\begin{figure}
    \centering
\centerline{\includegraphics[scale=0.6]{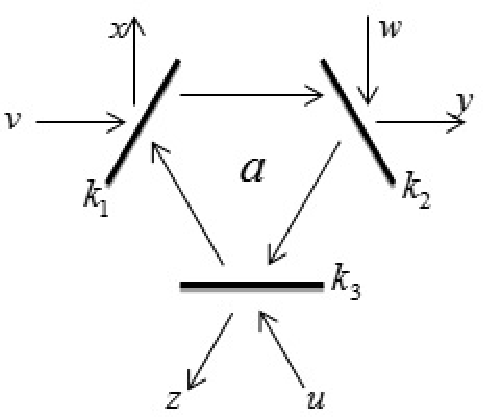}}
\caption{An uncertain cavity.\label{fig22}}
\end{figure}

In this section, an illustrative example is presented, which shows capabilities of the proposed decomposition algorithms and their consequences. A simpler version of this example has been considered in \cite{james2008h}. This example is an optical cavity coupled to three input channels $v,w,u$  , and three output channels $x,y,z$  as shown in figure \ref{fig22}. ${A_1}(t),{A_2}(t),{A_3}(t)$ represent the input fields of channels $v,w,u$, respectively. ${B_1}(t),{B_2}(t),{B_3}(t)$ also represent the output fields of channels $x,y,z$, respectively. Dynamics of this optical cavity is described by the annihilation, $a$, and creation, ${a^*}$, operators.

Let us define the input field $A(t) \triangleq {\left( {{A_1}(t),{A_2}(t),{A_3}(t)} \right)^T}$, the output field $B(t) \triangleq {\left( {{B_1}(t),{B_2}(t),{B_3}(t)} \right)^T}$, and the state vector $X(t) \triangleq \left( {a(t)} \right)$. The original uncertain plant parameters are:
\[\begin{gathered}
  L = [{C^ - },{C^ + }]\mathop X\limits^ \cup  (t) = \left( {\begin{array}{*{20}{c}}
  {\sqrt {{k_1} + \gamma } }&0 \\ 
  {\sqrt {{k_2}} }&0 \\ 
  {\sqrt {{k_3}} }&0 
\end{array}} \right)\mathop X\limits^ \cup  (t), \\ 
  S = {I_3},H = \delta {a^*}a, \\ 
\end{gathered} \]
where $\delta $ is the cavity frequency detuning. Also,  $\gamma $ denotes the uncertainty in the value of ${k_1}$. Prior knowledge about this uncertain parameters is taken into consideration as $\left| \delta  \right| \leqslant \bar \delta $ and $\left| \gamma  \right| \leqslant \bar \gamma $ .
\par Using the procedures described in Section \ref{sec}, the state-space realization is obtained for this system as follows:
\[d\mathop {\mathop X\limits^ \cup  }\limits^{}  = \left( {\begin{array}{*{20}{c}}
  { - \frac{{\lambda  + \gamma }}{2} - 2i\delta }&0 \\ 
  0&{ - \frac{{\lambda  + \gamma }}{2} + 2i\delta } 
\end{array}} \right)\mathop X\limits^ \cup  (t)dt + \left( {\begin{array}{*{20}{c}}
  { - \sqrt {{k_1} + \gamma } }&{ - \sqrt {{k_2}} }&{ - \sqrt {{k_3}} }&0&0&0 \\ 
  0&0&0&{ - \sqrt {{k_1} + \gamma } }&{ - \sqrt {{k_2}} }&{ - \sqrt {{k_3}} } 
\end{array}} \right)d\mathop A\limits^ \cup  (t),\]
\[d\mathop B\limits^ \cup  (t) = \left( {\begin{array}{*{20}{c}}
  {\sqrt {{k_1} + \gamma } }&0 \\ 
  {\sqrt {{k_2}} }&0 \\ 
  {\sqrt {{k_3}} }&0 \\ 
  0&{\sqrt {{k_1} + \gamma } } \\ 
  0&{\sqrt {{k_2}} } \\ 
  0&{\sqrt {{k_3}} } 
\end{array}} \right)\mathop X\limits^ \cup  (t)dt + d\mathop A\limits^ \cup  (t),\]
where $\lambda  = {k_1} + {k_2} + {k_3}$ is the nominal decay rate.
\par Separating uncertainties in SLH parameters in both additive and multiplicative manners leads to:

$$\Delta S = I,$$
\[\Delta L = \left( {\delta _c^ - ,\delta _c^ + } \right)\mathop {{X_\delta }}\limits^ \cup  (t) = \left( {\begin{array}{*{20}{c}}
  {\sqrt {{k_1}} (\sqrt {1 + \frac{\gamma }{{{k_1}}}}  - 1)}&0 \\ 
  0&0 \\ 
  0&0 
\end{array}} \right)\mathop {{X_\delta }}\limits^ \cup  (t),\]
$$\Delta H = \delta {a^*}a,$$
$${S_n} = I,$$
$${L_n} = \left( {{C_n}^ - ,{C_n}^ + } \right)\mathop {{X_n}}\limits^ \cup  (t) = \left( {\begin{array}{*{20}{c}}
  {\sqrt {{k_1}} }&0 \\ 
  {\sqrt {{k_2}} }&0 \\ 
  {\sqrt {{k_3}} }&0 
\end{array}} \right)\mathop {{X_n}}\limits^ \cup  (t),$$
$${H_n} = 0.$$
Subscripts $\delta $ and $n$  are used for the double-up state vector, $X(t)$, in order to distinguish between the states of the uncertain, ${X_\delta }(t) = \left( {{a_\delta }(t)} \right)$,  and nominal, ${X_n}(t) = \left( {{a_n}(t)} \right)$, subsystems. 
Using the decomposition algorithm proposed in Theorem \ref{th1}, the cavity is decomposed into the following subsystems:
$$\begin{gathered}
  \Delta  \equiv (I,\Delta L,\Delta H - \operatorname{Im} ({L_n}^\dag \Delta L)), \\ 
  {G_n} \equiv (I,{L_n},0), \\ 
\end{gathered} $$
where the state matrix of the nominal subsystems is:
$${\tilde A_n} = \left( {\begin{array}{*{20}{c}}
  { - \frac{\lambda }{2}}&0 \\ 
  0&{ - \frac{\lambda }{2}} 
\end{array}} \right),$$
which is Hurwitz stable since $\lambda  > 0$. Also, the state matrix of the uncertain subsystems is:
$${\tilde A_\delta } = \left( {\begin{array}{*{20}{c}}
  { - {k_1} - \frac{\gamma }{2} + \sqrt {{k_1}^2 + \gamma {k_1}}  - 2i\delta }&0 \\ 
  0&{ - {k_1} - \frac{\gamma }{2} + \sqrt {{k_1}^2 + \gamma {k_1}}  + 2i\delta } 
\end{array}} \right).$$
Matrices $\tilde A'$ and $\Delta \tilde A$ in (\ref{a234}) can be computed as:
\[\tilde A' = \left( {\begin{array}{*{20}{c}}
  {{k_1} - \sqrt {{k_1}^2 + \gamma {k_1}} }&0 \\ 
  0&{{k_1} - \sqrt {{k_1}^2 + \gamma {k_1}} } 
\end{array}} \right),\]
\[\Delta \tilde A = \left( {\begin{array}{*{20}{c}}
  { - \frac{\gamma }{2} - 2i\delta }&0 \\ 
  0&{ - \frac{\gamma }{2} + 2i\delta } 
\end{array}} \right).\]
It is obvious that:
$$\tilde A = {\tilde A_n} + \Delta \tilde A,$$
as it was expected. Thus, we have evaluated the additive perturbation to the state matrix. 
\par It is straightforward to check the validity of points \ref{a2} and \ref{a3} in Theorem \ref{th2}.
\par In order to perform robust stability analysis, the total decay rate is chosen to be $\lambda =k_1 +k_2 +k_3 =3$. Based on robust stability theorem \ref{Th33} (by numerical simulations, $\underset{\omega}{\inf} \underline{\sigma}(i\omega I- \tilde A_n )=1.5$), this uncertain cavity is guaranteed to be robustly stable if the norm of the additive perturbation $\Delta \tilde A$ does not exceed $\eta < 1.5$, or equivalently $\sqrt{{\frac{\Bar{\lambda}^2}{4}}+4\bar\delta}<1.5$. This example shows how the proposed theorems provide us with a systematic procedure to decompose an uncertain linear quantum network and facilitate robust stability analysis.	
\section{CONCLUSION AND FURTHER WORKS}\label{s4}
\par In this paper, a general uncertainty form for uncertain linear quantum input-output networks was considered, which was inspired by both natural properties and experimental aspects of such systems. Based on this uncertainty form, two decomposition theorems were developed. In the first theorem, the uncertain quantum network was decomposed into uncertain and nominal parts in cascaded connection, and in the second theorem, uncertainty decomposition was extended to state-space realizations of the sub-networks. Based on the proposed decomposition algorithm, as a potential application, a robust stability theorem for uncertain linear quantum networks was presented. Building on the proposed uncertainty decomposition scheme, authors' future work will focus on robust stability analysis and robust coherent controller design for uncertain linear quantum networks. Using methods introduced in \cite{azodi2017expectational,azodi2017stochastic}, a new stability analysis framework will be developed.


\section*{Acknowledgments}

\appendix
The following Lemmas are used in the proof of Theorems \ref{th1} and \ref{th2}:
\begin{lemma}\label{l1}
The following algebraic relations hold:
\begin{equation}
\begin{gathered}
  {(A + B)^\flat } = {A^\flat } + {B^\flat } \\ 
  \Delta {(A,B)^\flat } = \Delta ({A^\dag }, - {B^T}) \\ 
  \Delta (A + C,B + D) = \Delta (A,B) + \Delta (C,D) \\ 
  \Delta {({X^\dag }A,{X^\dag }B)^\flat }\Delta ({X^\dag }A,{X^\dag }B) = \Delta {(A,B)^\flat }\Delta (A,B) \\ 
\end{gathered} 
\end{equation}
\end{lemma}
\begin{proof}
Proofs are straight-forward.
\end{proof}
\begin{lemma}\label{l2}
The additional Hamiltonian term, $\operatorname{Im} \left( {L_2^\dag {S_2}{L_1}} \right)$, in (\ref{12}) corresponds to the double-up form $\tilde \Omega  = {\operatorname{Im} _\flat }(\tilde C_2^\flat {S_2}\tilde C_1^{})$ and induces $- i\tilde \Omega  =  - i{\operatorname{Im} _\flat }(\tilde C_2^\flat {S_2}\tilde C_1^{})$ to the state matrix of the  corresponding system (${\operatorname{Im} _{_\flat }}(X) \triangleq \frac{1}{{2i}}(X - {X^{_\flat }})$ ).
\end{lemma}
\begin{proof}
Refer to \cite{gough2010squeezing}.
\end{proof}

\nocite{*}
\bibliography{bib1}%

\clearpage

\end{document}